\documentclass[prl,aps,letterpaper,floatfix,superscriptaddress,preprintnumbers,twocolumn,10pt]{revtex4-1}

\usepackage{dcolumn}
\usepackage{bm}
\usepackage[dvips]{graphicx}
\usepackage{epsfig}
\usepackage{amsmath, amsfonts, amssymb, slashed}

\newcommand{\be}{\begin{equation}}
\newcommand{\ee}{\end{equation}}
\newcommand{\bee}{\begin{equation*}}
\newcommand{\eee}{\end{equation*}}

\usepackage{color}

\newcommand{\bea}{\begin{eqnarray}}  
\newcommand{\eea}{\end{eqnarray}}

\usepackage{feynmp}
\usepackage{tikz}
\begin{document}

\title{
See-Saw Composite Higgses at the LHC: \\
Linking Naturalness to the $750$ GeV Di-Photon Resonance}

\author{Jose Miguel No, Veronica Sanz and Jack Setford} 
\affiliation{Department of Physics and Astronomy, University of Sussex, 
Brighton BN1 9QH, UK}

\date{\today}

\begin{abstract}
We explore the possibility of explaining the recent $\sim 750$ GeV excesses observed by ATLAS and CMS in the $\gamma\gamma$ spectrum 
in the context of a compelling theory of Naturalness. The potential spin-zero resonance responsible for the excesses also requires the existence of new heavy charged states. We show that both such features are naturally realized in a see-saw Composite Higgs model for EWSB, where the new pseudo-Goldstone bosons 
are expected to be comparatively heavier than the SM Higgs, and the new fermions have masses in the TeV range. If confirmed, the existence of this new resonance could be the first stone in 
the construction of a new theory of Naturalness.
\end{abstract}

\maketitle


Composite Higgs models are a natural and attractive solution to the hierarchy problem of the Standard Model (SM), with the 
Higgs field realized as a pseudo-Goldstone boson of a spontaneously broken global symmetry $\mathcal{G} \to \mathcal{H}$. 
This  protects the Higgs mass  from being quadratically sensitive to high mass scales that could be present in the UV 
completion of the SM addressing key open questions, such as the nature of Dark Matter and the origin of the 
matter-antimatter asymmetry in the Universe. 

The global symmetry is however not exact, being explicitly broken {\it e.g.} by the gauge and Yukawa interactions of the Higgs doublet, which in turn yields 
a scalar potential for the Higgs and other pseudo-Goldstone bosons from the coset $\mathcal{G}/\mathcal{H}$. 
The number of pseudo-Goldstone scalars in the low energy theory and their $SU(2)_{\mathrm{L}} \times U(1)_{\mathrm{Y}}$ quantum numbers 
is directly related to the breaking and the embedding of the electroweak (EW) gauge group in $\mathcal{H}$~\cite{Agashe:2004rs,gripaios,Mrazek:2011iu,Bertuzzo:2012ya}.  

For these scenarios to be viable completions of the SM, the scalar potential must allow for EW symmetry breaking (EWSB), 
which requires a negative mass-squared term for the Higgs field. This is generically induced via fermionic 
loop contributions to the potential, coming dominantly from ``top partner" states. In order to 
achieve Natural EWSB together with a light Higgs mass these new fermions 
 cannot be much heavier than the EW scale, which creates a significant amount of tension as the current LHC limits push their mass towards the TeV scale.
 
In~\cite{Sanz:2015sua} an elegant solution to this problem was proposed, in the form of a see-saw type of EWSB: considering the 
sequential symmetry breaking pattern $SO(6) \to SO(5) \to SO(4)$ gives rise to a pseudo-Goldstone doublet $\phi$ and singlet $\eta$  from the first breaking, and another 
doublet $\theta$ from the second breaking. Due to the sequential pattern, the $\eta$, $\phi$ fields are expected to be significantly 
heavier than $\theta$, since $SO(6)/SO(5)$ breaking interactions would generate a mass for $\eta$, $\phi$ but not $\theta$. 
In this scenario, scalar potential terms of the form $\mu^2\,\phi^{\dagger}\theta + \mathrm{h.c.}$ would give rise to a mixing 
between the heavy and light scalar doublets, yielding after diagonalization a negative mass term for the light doublet eigenstate, which would trigger 
EWSB without the need of light top partners, those being now linked instead to the heavier pseudo-Goldstone scalars $\phi$, $\eta$. This greatly ameliorates the fine-tuning issues in these scenarios. Moreover, new states from strong dynamics contributing  to the potential of the heavy scalars could play a role in phenomenology.

Very recently, both ATLAS~\cite{ATLAS_diphoton} and CMS~\cite{CMS:2015dxe} collaborations have observed a prominent excess in the di-photon 
spectrum around $m_{\gamma\gamma} \sim 750$ GeV, which could be the first signature of new physics beyond the SM at the LHC 13 TeV Run, and indeed has attracted much attention~\cite{macro}. We take this excess seriously, albeit with a light heart, and identify it with one of the heavy scalars in the model, whose production and decays are induced by the heavy states. 
We cannot fail to note that the mass hierarchy between the two scalars is precisely what one would expect in this naturalness-motivated scenario.


\vspace{-3mm}

\section{The See-Saw Composite Higgs Model\label{SecCH654}}

Let us now discuss the main features of our setup (for a more detailed discussion, see~\cite{Sanz:2015sua}). 
The model features a global $SO(6)$ symmetry that is spontaneously broken via $SO(6) \rightarrow SO(5) \rightarrow SO(4)$.
The first breaking gives rise to five Goldstone bosons, a doublet under $SU(2)_L$, $\phi$ and a singlet $\eta$, while the second breaking gives rise to another 
doublet $\theta$. The presence of sources of explicit $SO(6)/SO(5)$ breaking in the UV theory may yield 
mass terms for the first set of Goldstone bosons
\begin{equation}
\label{LMass}
\mathcal L_\mathit{mass} = m_\phi^2 {\boldsymbol \phi}^2 + m_\eta^2 \eta^2,
\end{equation}
as well as mixing terms between the different sets of Goldstones 
\begin{eqnarray}
\label{LMixing}
\mathcal L_\mathit{mix} & = & A_1 F {\boldsymbol \phi} \cdot {\boldsymbol \theta} \frac{s_\theta}{|{\boldsymbol \theta}|} 
+ A_2 F \eta c_\theta + B_1 F^2 ({\boldsymbol \phi} \cdot {\boldsymbol \theta})^2 \frac{s_\theta^2}{|{\boldsymbol \theta}|^2}\nonumber \\
&+ &B_2 F^2 \eta^2 c_\theta^2 + 2B_3 F^2 \eta {\boldsymbol \phi} \cdot {\boldsymbol \theta} \frac{s_\theta c_\theta}{|{\boldsymbol \theta}|},
\end{eqnarray}
where ${\boldsymbol \phi} = (\phi^1 \, \phi^2 \, \phi^3 \, \phi^4)^T$ and ${\boldsymbol \theta} = (\theta^1 \, \theta^2 \, \theta^3 \, \theta^4)^T$
are vectors of $SO(4) \simeq SU(2)_L \times SU(2)_R$, the parameters $A_i$ and $B_i$ have mass dimension $[A] = 2$, $[B] = 0$, and
\begin{equation}
s_\theta = \sin\frac{|{\boldsymbol \theta}|}{F},\;\;c_\theta = \cos\frac{|{\boldsymbol \theta}|}{F}.
\end{equation}
%
In the $SO(5)$ invariant limit for (\ref{LMass}) and (\ref{LMixing}) we 
expect $m_\phi = m_\eta$, $A_1 = A_2$, $B_1 = B_2 = B_3$. We would expect deviations from this limit 
due to potential $SO(5)$ violating effects, expected to be comparable to the loop induced mass of the light doublet 
$\theta$, i.e. $\delta m^2 \sim \delta A \gtrsim m_\theta^2$.

It is worth emphasising that this model is free from many of the fine-tuning issues that plague ordinary Composite Higgs models. It is not difficult to achieve a satisfactory light Higgs potential (with a small physical Higgs mass and a \emph{vev} much below the breaking scale), with values of the $A, B$ coefficients that obey the above scaling that we expect. By effectively introducing a new scale the model also manages to relax the constraint that top partners need to be light.

Looking at (\ref{LMixing}), we first note that for $A_2 \neq 0$, the singlet field $\eta$ develops a {\it vev}, $\eta \to  \langle \eta \rangle + \eta$, with 
\begin{equation}
\langle \eta \rangle=-\frac{A_2 F}{2(m_\eta^2 + B_2 F^2)}
\label{etavev}
\end{equation}
At the same time, the term proportional to $A_1$ in (\ref{LMixing}) induces a mixing between the two doublets $\phi$, $\theta$ via the mass matrix 
\begin{equation}
\label{MassMatrix}
\begin{pmatrix} m_\phi^2 & \mu^2   \\ \mu^2 & m_{\theta}^2 - \frac{A_2}{2 F}\langle \eta \rangle - 
B_2 \langle \eta \rangle^2 \end{pmatrix}
\end{equation}
with $\mu^2 \equiv A_1/2 + B_3 F \langle \eta \rangle $. The mixing yields two (doublet) eigenstates $H$ and $h$, 
the latter being the light SM-like Higgs, which obtains a negative mass-squared for 
$\mu^2 > m_\phi \sqrt{m_{\theta}^2 - A_2\langle \eta \rangle/(2F) - B_2 \langle \eta \rangle^2}$.
The occurance of such a negative mass-squared term from the mixing of the two Goldstone $SU(2)_{\mathrm{L}}$ doublets, 
associated with the sequential global symmetry breaking pattern, is key in this framework, yielding viable EWSB \textit{\'a la} see-saw
(see~\cite{Galloway:2013dma,Chang:2014ida} for a similar realization of EWSB in other contexts). 
The rotation to the doublet mass eigenbasis is given by $\boldsymbol \phi = c_{\alpha} {\bf H} - s_{\alpha} {\bf h}$, 
$\boldsymbol \theta = s_{\alpha} {\bf H} + c_{\alpha} {\bf h}$, with $c_{\alpha} \equiv \cos \alpha$, $s_{\alpha} \equiv \sin \alpha$
and the rotation angle given by 
\begin{equation}
\tan 2\alpha = \frac{A_1 + 2B_3 F \langle\eta\rangle}{m_\phi^2 +\frac{A_2}{2 F} \langle\eta\rangle + B_2\langle\eta\rangle^2 - m_{\theta}^2}.
\end{equation}

From the discussion above it is clear that in the $SO(5)$ invariant limit for (\ref{LMixing}), a {\it vev} for the singlet field $\eta$ 
is needed for the see-saw EWSB mechanism to occur (since $A_1 = A_2$), while in the presence of a small $SO(5)$ breaking $\delta A$, 
$\langle\eta\rangle = 0$ could be a phenomenologically viable possibility. 

\vspace{2mm}

Expanding the Composite Higgs Scalar Potential (\ref{LMixing}) we find that the relevant terms involving $\eta$, $H$ and $h$ are
\begin{eqnarray}
\label{Pot_Expanded}
&-&\mu^2_h \,h^\dagger h + \mu^2_H \, H^\dagger H + (m_\eta^2 + B_2 F^2)\, \eta^2 \nonumber \\
&+& \left[ -\left( \frac{A_2}{2F} + 2 B_2 \langle\eta\rangle \right)c_\alpha^2 - 2B_3 F s_\alpha c_\alpha \right] \eta h^\dagger h \nonumber \\
&+& \left[ -\left( \frac{A_2}{2F} + 2 B_2 \langle\eta\rangle \right) s_\alpha^2 + 2B_3 F s_\alpha c_\alpha \right] \eta H^\dagger H  \nonumber \\
&+& \left[ -\left( \frac{A_2}{2F} + 2 B_2 \langle\eta\rangle \right)s_\alpha c_\alpha + 2B_3 F(c_\alpha^2 - s_\alpha^2) \right] \eta H^\dagger h + \mathrm{h.c.} \nonumber \\
&-&B_2 c_\alpha^2 \eta^2 h^\dagger h - B_2 s_\alpha^2 \eta^2 H^\dagger H  - 2B_2 s_\alpha c_\alpha \eta^2 H^\dagger h + \mathrm{h.c.}
\end{eqnarray}
with $\mu^2_h$ and $\mu^2_H$ the resulting squared-mass terms after the diagonalization of (\ref{MassMatrix}). 
 
\subsection{Couplings of $H$ and $\eta$ to $\gamma \gamma$}

\vspace{-3mm}

The couplings of the new heavy scalar states to $\gamma \gamma$ occur via loops of the heavy fermions responsible for the explicit
$SO(6)/SO(5)$ breaking (as discussed in detail in the Appendix). These generically transform both under $SU(3)_{\mathrm{C}}$ and $U(1)_{\mathrm{Y}}$, 
and as such may be responsible both for the production of these scalars at the LHC in gluon fusion $p p \,(g g) \to H,\eta$, and their subsequent
decay into $\gamma\gamma$. The effective couplings of $H,\eta = \varphi$ to SM gauge bosons are given by
\begin{equation}\label{LGauge}
\mathcal L_\mathit{G} = \frac{c_1}{4} \varphi \, B_{\mu\nu} B^{\mu\nu} + \frac{c_2}{4} \varphi\, W^i_{\mu\nu} W^{i\mu\nu} + \frac{c_3}{4} \varphi\, G^a_{\mu\nu} G^{a\mu\nu} \ , 
\end{equation}
with $i = 1,2,3$ and $a = 1,...,8$. The equation above is correct if we assumed $\eta$ is CP-even state~\cite{gripaios}, whereas if $\eta$ is a CP-odd particle, we would do the substitution of one of the field strengths by a dual, $B_{\mu\nu} B^{\mu\nu} \to B_{\mu\nu} \tilde B^{\mu\nu} \ldots$
 
We can re-express the above as
\begin{eqnarray}
\label{LGauge2}
\mathcal L_\mathit{G} &=& -\frac{g_{\gamma\gamma}}{4} \varphi \, F_{\mu\nu} F^{\mu\nu} - \frac{g_{z\gamma}}{4} \varphi\, F_{\mu\nu} Z^{\mu\nu} -
\frac{g_{zz}}{4} \varphi\, Z_{\mu\nu} Z^{\mu\nu} \nonumber \\
&-& \frac{g_{ww}}{4} \varphi\,  W_{\mu\nu} W^{\mu\nu} - \frac{g_{G}}{4} \varphi\, G^a_{\mu\nu} G^{a\mu\nu}
\label{EFT}
\end{eqnarray}
with $g_{\gamma\gamma} = c_1 \alpha_1 c^2_W + c_2 \alpha_2 s^2_W$, $g_{z\gamma} = (c_1 \alpha_1 - c_2 \alpha_2) s_{2W}$, 
$g_{zz} =  c_1 \alpha_1 s^2_W + c_2 \alpha_2 c^2_W$, $g_{ww} = 2 c_2 \alpha_2$, $g_{G} = c_3$, and $\alpha_{1,2}$ being
respectively $g^2/(4\pi)$ and $g'^2/(4\pi)$.
%
%
%
\vspace{2mm}

The relation between these effective operators and the heavy fermions depends on the fermion representation, their embedding in the SM and whether the scalars acquire a {\it vev}. In the Appendix we discuss in detail the embedding of the fermions in the electroweak gauge group. Among our fermionic bound states, all coloured fermions will participate in the coupling of the resonance to gluons in a universal fashion, and in the following we will denote their number as $N_{3}$, whereas the number of fermions contributing to the electroweak couplings will be denoted by $N_{ew}$. 

Focusing on fermions in a representation $\bf 4$ of $SO(5)$, the effective couplings of $\eta$ to electroweak gauge bosons are given by

\begin{table}[h]
\centering
\begin{tabular}{c | c  | c | c}
\hline
& $c_1$ & $c_2$& $g_{\eta \gamma\gamma}$ \\
\hline
$\langle \eta \rangle = 0 $ & $-  \frac{y_\eta N_{ew}}{2 M_{\Psi}}$ & $ \frac{y_\eta N_{ew} }{2 M_{\Psi}}$ & 0 \\
$\langle \eta \rangle \neq 0 $ & $ (\frac{1}{2} + 4\,X^2)   \,\frac{ -N_{ew} y_\eta^2\langle\eta\rangle}{M^2_{\Psi}}$ & $\frac{ -N_{ew} y_\eta^2}{2}\,\frac{\langle\eta\rangle}{M^2_{\Psi}}$ &  $-\frac{N_{ew} y_\eta^2 \langle\eta\rangle \alpha}{M_\Psi^2}(1+4X^2)$ \\
\hline
\end{tabular}
\label{tableeta}
\end{table}

where $M_{\Psi}$ is a common fermion mass, linked to  the strong dynamics responsible for the breaking of $SO(6)\to SO(5)$, hence $M_{\psi} \gg v$ naturally. For a non-vanishing $\langle \eta \rangle$, the heavy fermions get a correction to their mass term
\begin{equation}
y_\eta \langle \eta \rangle \overline{\Psi_4} \gamma_5 \Psi_4 = y_\eta   \langle \eta \rangle (\overline{\psi^1}\psi^1 + \overline{\psi^2}\psi^2-\overline{\psi^3}\psi^3 - \overline{\psi^4}\psi^4) \ ,
\end{equation}
which satisfies $y_\eta \langle \eta \rangle/M_\psi \ll 1$, see Eq.~\ref{etavev}. $X$ in this table denotes the $U(1)_X$ charge of $\Psi_4$, defined in the Appendix. 

An alternative possibility is for $\varphi = H$ to decay into two photons. We note that the eigenstate $\phi_4$ coupling to the heavy fermions does it as
\begin{equation}
y_\phi \phi_4 \overline\Psi_4 \gamma_4 \Psi_4 = y_\phi \phi_4 (\overline{\psi^1}\psi^3 + \overline{\psi^2}\psi^4) + \mathrm{h.c.}
\end{equation}
The mixing between $\phi$ and $\theta$ yields a correction to the fermion mass term after EWSB 
\begin{equation}
y_\phi s_\alpha v (\overline{\psi^1}\psi^3 + \overline{\psi^2}\psi^4) + \mathrm{h.c.},
\end{equation}
where $v = \langle h\rangle$ is the Higgs {\it vev}. In this case
\begin{equation}
\label{c1}
c_1 = (1/2 + 4\,X^2) y_\phi^2 \,\frac{s_\alpha v}{M^2_{\Psi}}, \quad c_2 = \frac{y_\phi^2}{2}\,\frac{s_\alpha v}{M^2_{\Psi}} \ , 
\end{equation}
and the coupling to photons is $g_{H\gamma\gamma}=  -\frac{N_{ew} y_h^2 s_\alpha v \alpha}{M_\Psi^2}(1+4X^2) $.

Finally, let us write the relation of  the branching ratio to photons respect to  other vector bosons, 
\bea
r_{XY} = \frac{BR(\phi\to XY)}{BR(\phi\to \gamma\gamma)}
\eea

Below we present the double ratios for $\eta$, although one would get similar expressions for $H$ with $y_\eta \langle \eta \rangle \to y_H v s_\alpha$. 

\begin{table}[h]
\centering
\begin{tabular}{c | c  | c | c}
\hline
& $r_{ZZ}$ & $r_{Z\gamma}$ & $r_{WW}$ \\
\hline
$\eta$ & $\frac{2.718}{(1+4X^2)^2} r_{\eta} $ & $\frac{1.9}{(1+4X^2)^2} r_\eta$ & $\frac{21.11}{(1+4X^2)^2} r_\eta$ \\
$H$ & $\frac{3.682 + 4.356 X^2 + 1.289 X^4}{(1+4X^2)^2}$ & $\frac{5.917 + 20.77X^2 + 18.24X^4}{(1+4X^2)^2}$ & $\frac{21.11}{(1+4X^2)^2}$ \\
\hline
\end{tabular}
\label{tableBRs}
\end{table}
where $r_\eta=\left(\frac{M_\Psi}{y_\eta \langle\eta\rangle}\right)^2$
For a canonical example of $X=\pm 1/2$, the ratios for the heavy Higgs are 1.2, 3.1 and 5.3, respectively, whereas for the $\eta$ particle they are a function of the fermion masses and couplings, namely $r_\eta \times$ (0.7,0.5,5.3), respectively. Given these branching ratios, Run2 LHC may be able to explore decays of the resonance to other states $WW$, $ZZ$ and $Z\gamma$~\cite{johnnew}. Moreover, current limits on these branching ratios from Run1 can be read in Ref.~\cite{johnnew}, leading to a limit on $r_\eta \lesssim 10$ from heavy Higgses decaying to $WW$. Therefore, the mass splitting among the fermions cannot be too small, typically of order $\sim M_\psi/3$.
 
\section{Di-Photon Signatures at the LHC}

Using the results from the previous sections, we now analyze the possibility that either $H$ or $\eta$ in our framework correspond to the potential 
di-photon resonance observed by both ATLAS and CMS around $m_{\phi} \sim 750$ GeV. 

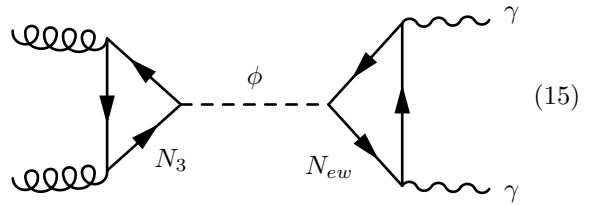
\begin{fmffile}{production_decay}
\begin{equation}
\label{production_decay}
\begin{tikzpicture}[baseline=(current bounding box.center)]
\node{
\fmfframe(1,1)(1,1){
\begin{fmfgraph*}(150,80)
\fmfleft{v1,v2}
\fmfright{v3,v4}
\fmflabel{$\gamma$}{v3}
\fmflabel{$\gamma$}{v4}
\fmf{curly,tension=.05}{v1,i1}
\fmf{curly,tension=.05}{v2,i2}
\fmf{plain_arrow,tension=.01}{i2,i1}
\fmf{plain_arrow,tension=.01,label=$N_3$,l.side=right}{i1,i3}
\fmf{plain_arrow,tension=.01}{i3,i2}
\fmf{dashes,tension=.01,label=$\phi$,l.side=left}{i3,i4}
\fmf{plain_arrow,tension=.01,label=$N_{ew}$,l.side=right}{i4,i5}
\fmf{plain_arrow,tension=.01}{i5,i6}
\fmf{plain_arrow,tension=.01}{i6,i4}
\fmf{wiggly,tension=.1}{i5,v3}
\fmf{wiggly,tension=.1}{i6,v4}
\fmffreeze
\fmfshift{-30,12}{v1}
\fmfshift{-30,-12}{v2}
\fmfshift{30,8}{v3}
\fmfshift{30,-8}{v4}
\end{fmfgraph*}
}
};
\end{tikzpicture}
\end{equation}
\end{fmffile}

\vspace{2mm}

We first note that in order for any new scalar $\varphi$ to have a sizeable branching fraction into $\gamma\gamma$, its tree-level decays into 
SM particles must be absent or heavily suppressed (see {\it e.g.} the discussion in~\cite{}). Then, for $\varphi = \eta$, the term 
$\eta h^\dagger h$ in (\ref{Pot_Expanded}) poses a potentially important obstacle towards achieving a sizeable $\mathrm{Br}(\eta \to \gamma\gamma)$. The partial width
$\Gamma(\eta\to h h)$ is given by 
\begin{equation}
\Gamma(\eta\to h h) = \frac{\kappa_{\eta hh}^2}{8\pi m_S} \sqrt{1- \frac{4\,m^2_h}{m_\eta^2}}
\end{equation}
with 
\begin{equation}
\kappa_{\eta hh} \equiv \left[ -\left( \frac{A_2}{2F} + 2 B_2 \langle\eta\rangle \right)c_\alpha^2 - 2B_3 F s_\alpha c_\alpha \right]\, ,
\end{equation}
such the relation $\Gamma(\eta\to h h) \gg \Gamma(\eta\to \gamma\gamma)$, would lead to
\begin{equation}
\kappa_{\eta hh} \lesssim   \frac{Q^4_{\Psi}\, y^2_\eta\, \alpha^2_{\mathrm{EW}}}{4 \pi^2}\, \frac{m_\eta^2}{M_{\Psi}}
\end{equation}
with $Q_{\Psi}$ the charge of the heavy fermions running in the $\eta\to \gamma\gamma$ loop.
Moreover, after EWSB $\kappa_{\eta hh} \neq 0$ gives rise to singlet-doublet mixing, such that the singlet-like mass eigenstate inherits a small amount of the Higgs couplings to 
SM particles. While the value of this mixing $\beta$ is constrained by a combination of LHC measurements of Higgs signal strengths and EW precision observables to 
$\sin(\beta) <  0.32$ at 95 \% C.L. for $m_\eta \sim 750$ GeV~\cite{Gorbahn:2015gxa}, admixtures below this value may still yield 
$\Gamma(\eta\to WW,ZZ,t\bar{t}) \gg \Gamma(\eta\to \gamma\gamma)$.

\vspace{2mm}

For $\varphi = H$, this potential problem does not arise since $H$ and $h$ do not mix (by construction), and $H$ does not have a priori 
any dangerous tree-level decays into SM particles.   

The signal strength compatible with the resonance in diphotons corresponds to a best-fit value for the total cross-section of~\cite{johnnew}
\bea
 \sigma(p p \to \phi \to \gamma\gamma ) = 6.2 \pm 1.0 \text{fb}  \ ,
\eea
and one can relate in this scenario this value to the fermion parameters responsible for the production and decay. First, it can be easily calculated in the narrow width approximation as
\bea
\sigma(p p \to \phi \to \gamma\gamma ) = \sigma_{prod} (g g \to \phi ) . BR(\phi \to \gamma \gamma)
\eea

In terms of the effective Lagrangian Eq.~\ref{EFT} these read
\bea
\Gamma(\phi \to g g) =  \frac{g_{G}^2}{8 \pi} \, m_\phi^3 \ , \hspace{.1cm } \Gamma(\phi \to \gamma \gamma) =  \frac{g_{\gamma\gamma}^2}{64 \pi} \, m_\phi^3 \ . 
\eea

Bounds on these couplings can be obtained from Fig. 3 in Ref.~\cite{johnnew}, resulting in a bound of $1/g_{\gamma\gamma}< 50$ TeV, which then translates in a bound on the fermion parameters $M_{\psi} \simeq N_{ew} (0.1-0.2)$ TeV, where we used the constraint from the branching ratio of $WW$ mentioned before. Direct searches of vector-like fermions as well as the constraint that $M_\psi> m_\phi$ leads to a rough estimate on the number of electroweak degrees of freedom contributing to the diphoton coupling, namely $N_{ew} \simeq {\cal O}(4-10)$.

\vspace{-3mm}

\section{Astrophysical and cosmological consequences}

There are a number of cosmological and astrophysical consequences of this scenario which deserve a more detailed study. 

Let us discuss first Dark Matter (DM). The neutral heavy fermions in our model can play the role of DM, and the resonance $\phi$ can be the mediator. Similar scenarios have been discussed in the literature in the context of radion/dilaton and axion mediators, for the CP-even~\cite{radion-DM} and CP-odd~\cite{axionDM} cases. This means that in our case the CP-odd $\eta$ must be the mediator, since then the annihilation cross section is s-wave unsuppressed. In this context a simple choice is then $X=\pm 1/2$ in Eq.~\ref{Xchoices} for one fermion multiplet, which leads to two neutral fermions. Among these, the lightest one will be DM, with a small splitting with the next state of order $(y_\phi v s_\alpha/M_\psi)^2$ or $(y_\eta \langle \eta \rangle/M_\psi)^2$. This will lead to a model similar to inelastic DM~\cite{iDM} or pseudo-Dirac DM~\cite{pDDM}, with coannihilations playing an important role. The main annihilation process would be to gluons, as $\phi$ decays predominantly to gluons. The relic abundance is then proportional to a combination $4 \pi^3 M_\psi^2/(y_\eta \alpha_s)^2$. Values in the range $M_\psi \sim$ TeV and $y_\eta \lesssim {\cal O}(1)$ lead to successful relic abundance. 

DM in this scenario would also produce $\gamma$-rays via the coupling of $\phi$ to $\gamma \gamma$ and $Z\gamma$. Understanding the correlations of the 750 GeV signal with possible lines in the spectrum measured by Fermi-LAT and HESS~\cite{Fermi-hess} could lead to a selection of fermionic representations in this model.

Finally, there is a tantalizing correlation between DM and Baryogenesis in this model. As we mentioned before, efficient annihilation requires a pseudo-scalar $\eta$ mediator, and the generation of the diphoton signal implies $\eta$ would get a {\it vev}, hence breaking spontaneously CP.  Additional fermionic states, new scalars and CP-breaking are excellent starting points to explore electroweak baryogenesis~\cite{EWPT-review} in this model.

\section{Summary}

In this paper we have presented an explanation of the diphoton signal seen by ATLAS and CMS in terms of a fully natural Composite Higgs model. The model features a new spectrum of composite scalars, with masses of order TeV. We find that these new states can decay via loops of vector-like heavy fermions and reproduce the observed diphoton excess.
The mass hierarchy between the Higgs-like doublet and the new scalars is a crucial and natural feature of the see-saw Composite Higgs model, and thus the new states are completely natural components of the model.

We have also identified a Dark Matter candidate: with a suitable $U(1)_X$ charge assignment the vector-like fermions can form neutral states that will behave as inelastic / pseudo-Dirac DM. For natural values of the model parameters we find that the model leads to successful relic abundance.

Arriving at a satisfying solution to the hierarchy problem without resorting to fine-tuning is a long standing challenge. Most potential solutions to the problem lead to us to expect new resonances around the TeV scale. If the recent diphoton signal is the first such observation, we believe the model we have presented succeeds in explaining the data in a coherent, and most importantly natural, fashion. 
\vspace{-3mm}

\begin{center}
\textbf{Acknowledgements} 
\end{center}

\vspace{-2mm}

This work is supported  by  the  Science  Technology  and  Facilities  Council  (STFC)  under  grant  number
ST/L000504/1. J.M.N. is supported by the People Programme (Marie curie Actions) of the European Union Seventh 
Framework Programme (FP7/2007-2013) under REA grant agreement PIEF-GA-2013-625809.

\section{Appendix: Heavy Vector-Like Fermions \& $SO(6)/SO(5)$ Breaking}

\vspace{-3mm}

Let us consider possible sources of $SO(6)/SO(5)$ breaking by introducing vector like heavy fermions in 
some representation of $SO(5)$ and allowing them to have Yukawa couplings to the heavy set of Goldstone bosons.
Below we discuss two different possibilities.

\vspace{-3mm}

\subsection{I. Spinor Representation}

\vspace{-3mm}

One simple possibility is for the heavy fermions $\Psi$ to be in the spinor ${\bf 4}$ representation of $SO(5)$. 
To calculate their contribution to the heavy Goldstone potential, we must embed the ${\bf 4}$ of $SO(5)$ in the spinor ${\bf 8}$ of $SO(6)$.
Assuming only one multiplet of $SO(5)$, $\Psi_4$, there are two possible embeddings into spinors of $SO(6)$
\begin{equation}
\Psi_+ = {\Psi_4 \choose 0},\;\;\Psi_- = {0 \choose \Psi_4}.
\end{equation}
The $SO(6)$ invariant effective Lagrangian is
\begin{eqnarray}
\label{L_eff}
\mathcal L_\mathit{eff} &=& \Pi_0^+(p) \overline\Psi_+ \slashed 
p \Psi_+ + \Pi_0^-(p) 
\overline\Psi_- \slashed 
p \Psi_- \nonumber \\ 
&+& \Pi_1(p) \overline\Psi_+ \Gamma_6^i \Sigma_i \Psi_- + \mathrm{h.c.}\, ,
\end{eqnarray}
where $\Gamma_6$ are the Gamma matrices of $SO(6)$ and
\begin{equation}
\Sigma = F \frac{\sin (\tilde \phi/F)}{\tilde \phi} (\phi^1, \phi^2, \phi^3, \phi^4, \eta, \tilde \phi \cot(\tilde \phi / F))
\end{equation}
with $\tilde \phi^2 = \boldsymbol \phi^2 + \eta^2$. This directly yields a $\sin^2$ Coleman-Weinberg potential
\begin{equation}
V(\phi,\eta) = - \bar{\alpha} \sin^2(\tilde \phi/F),
\end{equation}
where $\bar{\alpha}$ is an integral over the various form factors:
\begin{equation}
\bar{\alpha} = 4N_c \int \frac{d^4 p_E}{(2\pi)^4} \frac{F^2 \,\Pi_1^2(p_E)}{p_E^2 \,[\Pi_0^+(p_E) + \Pi_0^-(p_E)]^2}.
\end{equation}

The Yukawa couplings to the heavy fermions are given by the low-energy expression:
\begin{equation}
y_{\eta,\phi} = \frac{\Pi_1(0)}{\Pi_0^+(0) + \Pi_0^-(0)}.
\end{equation}

Next, we need to obtain the $U(1)_{\mathrm{Y}}$ charges of the heavy vector-like fermions. We take the global symmetry to be $SO(6) \times U(1)_X$, 
broken to $SO(4) \times U(1)_X$, in which the SM gauge group $SU(2)_L \times U(1)_Y$ is embedded. We then have $Y = T^{3R} + X$. $\Sigma$ is uncharged under $U(1)_X$. 
Therefore any charge asignment is allowed for $\Psi_4$ leaving \eqref{L_eff} invariant under the symmetry: if $\Psi_4$ does not contain any SM fermions then we 
have no constraints on the $U(1)_X$ assignments. Let $X$ be the charge of the $\Psi_4$ under $U(1)_X$. 
Then the charges of the component fermions under $U(1)_Y$ are:
\begin{equation}
\Psi_4 = \begin{pmatrix} \psi^1 \\ \psi^2 \\ \psi^3 \\ \psi^4 \end{pmatrix} \rightarrow \begin{pmatrix} X \\ X \\ X + 1/2 \\ X - 1/2 \end{pmatrix}.
\label{Xchoices}
\end{equation}
The quantum numbers under $T_{3L}$ are
\begin{equation}
\begin{pmatrix} 1/2 \\ -1/2 \\ 0 \\ 0 \end{pmatrix}.
\end{equation}

\subsection{II. Different Representations}

There is another possibility, namely that the $SO(6)/SO(5)$ Goldstone bosons couple to heavy fermions in different representations of $SO(5)$, for instance:
\begin{equation}
(\overline \Psi_5 \cdot \Sigma) \Psi_1,
\end{equation}
where $\Psi_5$ and $\Psi_1$ are a vector and a singlet under $SO(5)$, respectively. When the light Higgs gets a {\it vev}, 
another mass insertion on any of the fermion propagators can close the loop. Alternatively $SO(5)$ violating effects might interpolate an 
$SO(5)$ breaking mass term of the form $m \overline\Psi_5 \Psi_1$, where $m$ is expected to be of comparable order to $m_\theta$, since they are both 
induced by $SO(5)$ violating effects. In these cases we can perhaps get coefficients $c_{1,2}$ scaling as $\sim y^2 m$, possibly avoiding the factor of 
$s_\alpha$ in \eqref{c1}.

A further possibility would be a Yukawa like coupling of the form $\overline\Psi_{10} \Sigma \Psi_5$, having the invariant structure ${\bf\overline{10}\;5\;5 }$.



\end{document}